\documentclass[aps,prl,twocolumn,showpacs,preprintnumbers,amsmath,amssymb]{revtex4}
\usepackage{graphicx}
\usepackage{dcolumn}
\usepackage{bm}

\begin{document}
\title{Reconstructing supernova-neutrino spectra using low-energy beta-beams}
\author{N.~Jachowicz, }
\email{natalie.jachowicz@UGent.be}
\author{G.C.~McLaughlin}
\email{Gail_McLaughlin@ncsu.edu}
\affiliation{Department of Subatomic and Radiation Physics, Ghent University, Proeftuinstraat 86,  B-9000 Gent, Belgium.}
 \affiliation{Department of Physics, North Carolina State University, 
Raleigh, North Carolina 27695-8202.}
  
\date{\today}

\begin{abstract}Only weakly interacting, neutrinos are the principal messengers  reaching us from the center of a supernova.
Terrestrial neutrino telescopes, such as SNO 
 and SuperKamiokande, can provide 
precious information about the processes in the core of a collapsing and 
exploding star. 
But the information about the supernova that a neutrino 
detector 
can supply, is restricted by the fact that little experimental data on the neutrino-nucleus 
cross sections exists and by the uncertainties in theoretical calculations. In this letter, we propose a novel procedure that determines 
the response of a target nucleus in a 
supernova-neutrino detector 
directly, by using low-energy beta-beams. 
We show that fitting  'synthetic' spectra, constructed by taking linear combinations of beta-beam spectra, to the original supernova-neutrino spectra reproduces the folded differential cross sections very accurately.
Comparing the response in a terrestrial detector to these synthetic responses  
provides a direct way to determine
the main parameters of the supernova-neutrino energy-distribution. 
\end{abstract}
\maketitle

Understanding the  mechanism of a
 Type-II 
 core-collapse supernova is a longstanding problem. While the general nature of these events is understood
to be the explosion that results from the collapse of a massive star, a 
completely self-consistent model   of the explosion
does not yet exist.
Weak physics and  neutrino-interactions are widely regarded as 
central players in the explosion mechanism.  The core is so hot and dense
that even the neutrinos are trapped. These
neutrinos scatter out of the core on a timescale of about ten seconds,
taking away with them the vast majority of the gravitational binding energy 
released in the collapse. 
The handful of neutrinos that      were  detected from Supernova 1987A confirmed this 
basic picture of the nature of core-collapse supernovae. 

 The development of neutrino facilities and telescopes during the last 
decades  ensures
that a similar event today would  provide much more detailed information. 
When the next galactic supernova occurs,
neutrino detectors will be on the watch, using 
different targets and detection mechanisms  to catch neutrinos of various 
energies and flavors, thus gathering information about supernova (SN) neutrinos  
and about the supernova processes that generated and released them. 
Since the neutrinos scatter last at the neutrino sphere within the stellar
core, their energy spectrum encodes precious information about the
center of these astrophysical phenomena.

The extraction of the information SN neutrinos are carrying, 
however faces 
several intertwined problems.  
Uncertainties  in the description of neutrino scattering with nuclear matter,
general relativistic effects, and the effect of rotation 
 are carried over into the related production and release of neutrinos, and 
induce uncertainties in the energy spectra and flavor distributions of the 
emitted particles \cite{Liebendoerfer:2003es,Pons:2001ar,burrows}.
Neutrino oscillations will
 mix the original spectra on their way 
out of the exploding star, thereby veiling the information they can provide
\cite{Engel:2002hg,Fuller:1998kb,tak}. Beyond doubt, unraveling the information SN neutrinos deposit  in a detector
involves the need for an optimization of the 
interpretation of the signal.

 Neutrino-nucleus scattering is
an important detection mechanism.  The
cross sections are relatively large
 and reactions occur 
in the charged (CC) as well as
neutral-current (NC) channels.
The main nuclei of interest are  
$^{16}$O in SuperKamiokande \cite{SK}, the deuteron for SNO \cite{SNO}, $^{56}$Fe in LVD \cite{LVD}, $^{56}$Fe   and $^{208}$Pb 
for the proposed  LAND \cite{land} and OMNIS \cite{omnis} detectors and $^{12}$C for KamLAND \cite{vogel}, BooNe \cite{miniboone}, and LENA \cite{lena}.

The extent to which we can determine the physics of the supernova
from what we see in the detector depends on our understanding of the
neutrino-nucleus cross sections. 
\begin{figure}
\vspace*{5.8cm}
\special{hscale=33 vscale=33 hsize=1500 vsize=600
         hoffset=-15 voffset=183 angle=-90 psfile="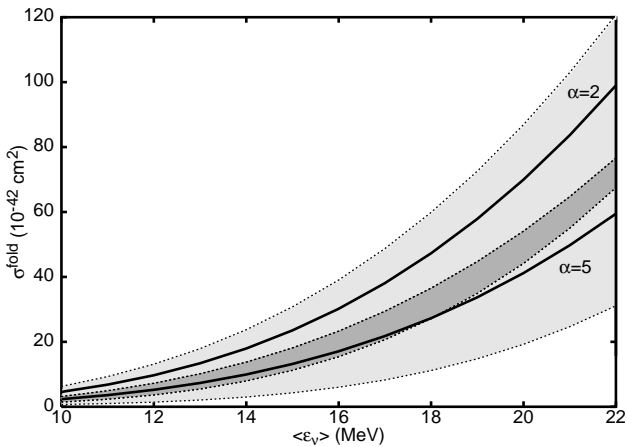"}
\caption{Uncertainty on theoretical cross sections for NC neutrino-scattering off $^{208}$Pb, folded with  different PL spectra, characterized by average energy $\langle\varepsilon_{\nu}\rangle$ and width parameter $\alpha$ \cite{keil}. The cross sections were taken from Refs.~\cite{kolbe,ikkepb,Engel:2002hg}. The full line represents the average of these predictions, the borders of the shaded regions are determined by the largest and smallest values of the results \cite{kolbe,ikkepb,Engel:2002hg}.}
\label{ogam}
\end{figure} 
However, the available  experimental neutrino-nucleus scattering data in the 
relevant energy range is limited.  Neutrino cross sections are small 
and monochromatic neutrino beams are not available. 
The shape of the neutrino  energy-distribution experimentally in use thus far \cite{karmen,nusns}, while in the same energy range as that of the SN neutrinos,
is      not the same as
an SN$\nu$ spectrum. 
Moreover, theoretical estimates of cross sections and response details
at these low energies are largely subject to nuclear structure uncertainties and model dependencies. Fig.~\ref{ogam} shows that the uncertainties in $^{208}Pb(\nu,\nu')$ cross section calculations of $^{208}Pb(\nu,\nu')$  induce large deviations in the corresponding folded SN$\nu$ responses. The overlap of the shaded regions in the figure make clear how uncertainties in cross sections calculations will impede the reconstruction of the original spectrum's parameters from an SN$\nu$ signal.
It is 
these problems
that we want to address using the strength and flexibility of the low-energy beta-beam concept \cite{volpe1}.
Low-energy beta-beam facilities would provide intense and well-characterized neutrino beams with energies typically in the 10 to 100 MeV range \cite{volpe2,cernbeam}.
In this letter,
we show that  taking appropriate combinations of beta-beam  neutrino cross-sections allows one to reconstruct differential SN$\nu$ cross-sections, and that this technique provides an efficient way to reconstruct the SN$\nu$ spectrum from the signal it induced in a detector.
This solves two problems : first, we avoid uncertainties and model dependencies rising in theoretical cross-section predictions. Second,  we address the complications raised by the lack of monochromatic neutrino beams. 

Since (anti)neutrinos of different energies and flavors decouple at different points in the core, the neutrino spectrum $n_{SN}(\varepsilon_{\nu})$
is not a purely thermal one. 
Common parameterizations are given by a Fermi-Dirac (FD) $n_{FD}(\varepsilon_{\nu})=\frac{N}{T^3}\frac{\varepsilon_{\nu}^2}{1+e^{\varepsilon_{\nu}/T-\eta}}$ or a power-law (PL) spectrum $n_{PL}(\varepsilon_{\nu})=\left(\frac{\varepsilon_{\nu}}{\langle \varepsilon_{\nu}\rangle}\right)^{\alpha} e^{-\frac{\left(\alpha+1\right)\varepsilon_{\nu}}{\langle\varepsilon_{\nu}\rangle}}$\cite{keil}. Both parameterizations yield similar energy distributions, characterized by the average energy or temperature $T$ of the neutrinos and the width of the spectrum.
The average energies $\langle\varepsilon_{\nu}\rangle$  vary roughly between 12 and 18 MeV for $\nu_e$ and $\overline{\nu}_e$'s, and between 16 and 22 MeV for $\nu_{\mu}$ and $\nu_{\tau}$. 
The width of the spectrum characterized by the parameters $\eta$ or $\alpha$ influences the weight of its high-energy tail \cite{keil}.
 
Similar to the shape of SN$\nu$ spectra, beta-beam spectra  are characterized by long tails, while their average energy and precise shape depend on the boost-factor $\gamma$ of the primary beam and the geometry of the experimental setup \cite{volpe1,McLaughlin:2004va, volpe2}.  The folded cross sections for 
neutrino-scattering processes show a very smooth behavior as a function of the $\gamma$-factor of the beta-beam spectrum \cite{volpe2}. Hence, a limited number of measurements can be used to infer the full $\gamma$-dependence  of the folded cross section. 
Here, we  suggest representing the SN$\nu$ spectra $n_{SN}$ as  linear combinations of  beta-beam spectra
$n^{\gamma}(\varepsilon_{\nu})$ \cite{McLaughlin:2004va}, by
\begin{equation}
n_{fit}({\varepsilon_{\nu}})=\sum_{i=1}^N a^{\gamma_i} n^{\gamma_i}(\varepsilon_{\nu}).
\label{const}
\end{equation}

The energy 
transfer to the target nucleus and the excited state $\omega$ it is left in, determine the 
decay mode of the nucleus.  Therefore,
the reaction products that will be observed in the detector
are fully determined by the differential folded response. 
In order to keep the discussion as general as possible, we do not restrict our  study to e.g. electron emission or nucleon knockout,  but we will use full differential cross sections to illustrate the proposed procedure. 

 The  detector response to a beta-beam neutrino spectrum is then given by 
$\int d\varepsilon_{\nu} \, \sigma(\varepsilon_{\nu},\omega) \, n^{\gamma_i}(\varepsilon_{\nu})$.
An actual supernova will produce a neutrino response determined by
$
\sigma_{signal}^{fold}(\omega)= \int_0^{\infty} d\varepsilon_{\nu}\sigma(\varepsilon_{\nu},\,\omega)\,n_{signal}(\varepsilon_{\nu})$. Adopting the philosophy of Eq.~\ref{const}, this quantity can be represented as a linear combination of beta-beam responses~:
\begin{equation}
\sigma^{fold}_{fit}(\omega) =\sum_{i=1}^N a^{\gamma_i} \int d\varepsilon_{\nu} \, \sigma(\varepsilon_{\nu},\omega) \, n^{\gamma_i}(\varepsilon_{\nu}) \label{linresp}.
\end{equation}
 Once a measured SN response  has been obtained, one can find the expansion parameters 
$a^{\gamma_i}, \gamma_i$ by fitting the  neutrino signal to the combinations of Eq.~\ref{linresp}. 
In the remainder of the paper, we first show that only a few beta-beam spectra in the linear combination (\ref{linresp}) are sufficient for an accurate reconstruction  of the SN$\nu$ signal,  then we illustrate that using this technique it is possible to determine the neutrino spectrum from the supernova without knowledge of $\sigma(\epsilon,\omega)$.

\begin{figure}
\vspace*{5.85cm}
\special{hscale=33 vscale=33 hsize=1500 vsize=600
         hoffset=-15 voffset=183 angle=-90 psfile="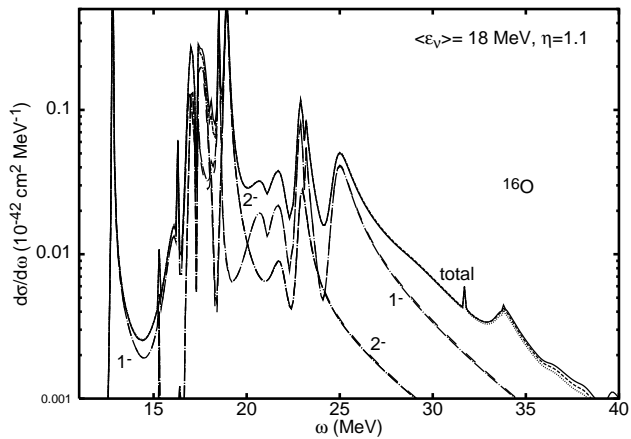"}
\caption{Comparison between differential cross section for NC scattering on $^{16}$O \cite{ikkepb}, folded with an FD distribution with  average energy  18 MeV and width, characterized by the degeneracy parameter, of $\eta$=1.1 \cite{keil},  and synthetic spectra with N=3 (dotted) and N=5 components (dashed). The choice of $\gamma$ was restricted to integer values between 5 and 15. The major multipole contributions (long-dashed) are compared with results of the $N=5$ fit (dashed-dotted).} 
\label{odif}
\end{figure}

The reconstruction is realized
by minimizing the expression
$\int d\varepsilon \, |n_{fit}(\varepsilon_{\nu})-n_{SN}(\varepsilon_{\nu})|$, varying the expansion coefficients $a^{\gamma_i}$ and boost factors $\gamma_i$.
In Fig.~\ref{odif} we show the corresponding differential folded cross sections for an $^{16}$O target and  a typical SN$\nu$ spectrum.
It is clear that the total strength as well as the position and width of the resonances are reproduced in a very convincing way, for a very small number $N$ of  beta-beam spectra in the fit.
The method is equally successful in reproducing the separate multipole contributions. This provides  evidence for the usefulness of  the proposed method  in predicting the yield in an SN$\nu$ detector, without going through the intermediate step of using a nuclear-structure calculation.
\begin{figure}
\vspace*{5.5cm}
\special{hscale=33 vscale=33 hsize=1500 vsize=600
         hoffset=-15 voffset=183 angle=-90 psfile="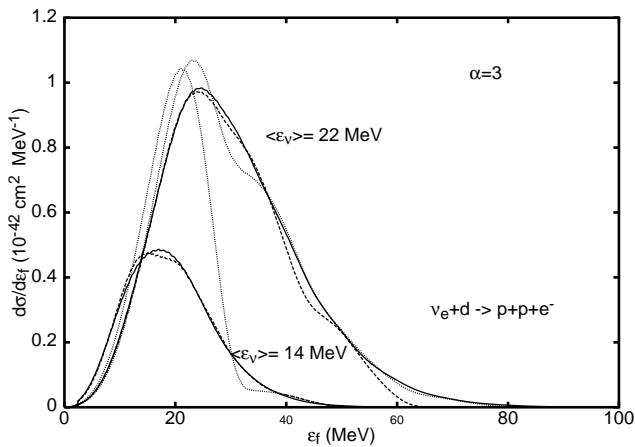"}
\caption{Comparison between differential cross sections \cite{kubo} for the CC reaction $\nu_e+d\rightarrow p+p+e^-$, folded with a PL SN$\nu$ spectrum (full line) and synthetic spectra with 5 components, using $\gamma$-factors ranging from 5 to 15 (dotted) and including lower energy beta-beam spectra with $\gamma$=3 to 15 (dashed).
}
\label{deutdif2}
\end{figure} 
The deuteron has a lower threshold than 'massive' nuclei, and cross sections peak at substantially lower energies.  
This has as  consequence that constructed 
spectra
must contain a beta-beam component at fairly low $\gamma$.
Fig.~\ref{deutdif2} shows that the fit
works well in reproducing the position and strength of the resonances. 
For spectra with smaller average energies the agreement is improving considerably when including lower gammas in the fit. 
\begin{figure*}
\vspace*{5.8cm}
\special{hscale=33 vscale=33 hsize=1500 vsize=600
         hoffset=-15 voffset=183 angle=-90 psfile="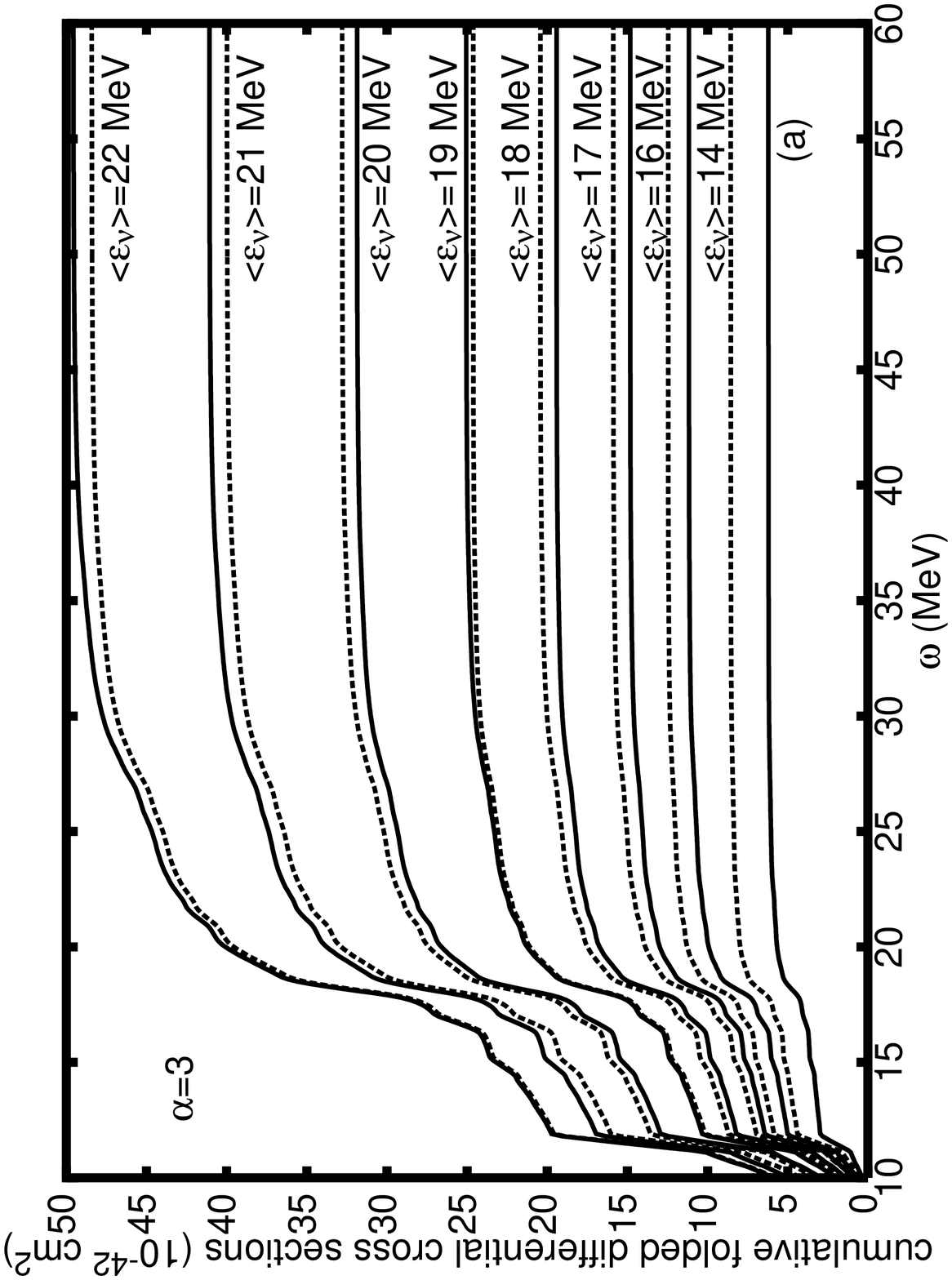"}
\special{hscale=33 vscale=33 hsize=1500 vsize=600
         hoffset=235 voffset=183 angle=-90 psfile="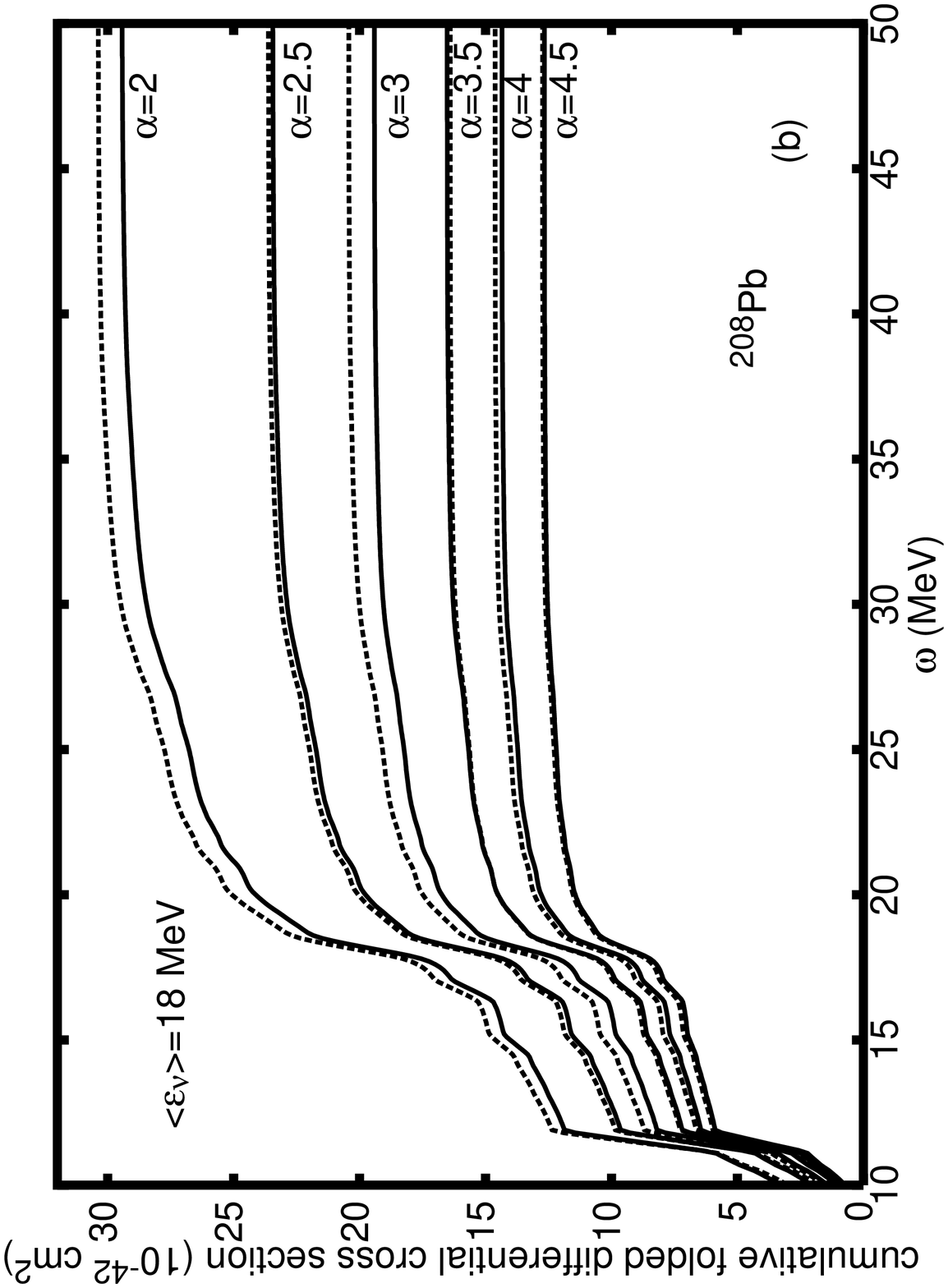"}
\caption{Comparison between cumulative cross sections for NC scattering on $^{208}$Pb, folded with a PL SN$\nu$ spectrum (full line) and synthetic spectra with 5 components (dashed), for different average energies (left) and widths (right panel).}
\label{pbdif}
\end{figure*}
 \begin{figure}
\vspace*{5.5cm}
\special{hscale=33 vscale=33 hsize=1500 vsize=600
         hoffset=-15 voffset=183 angle=-90 psfile="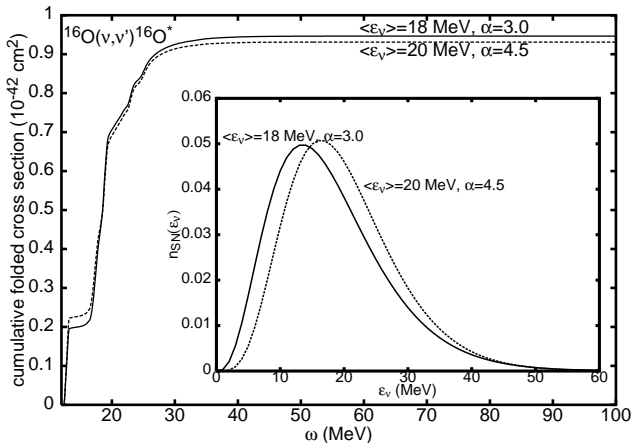"}
\caption{Comparison between PL spectra and the cumulative cross sections for selected spectrum parameters.}
\label{cumgelijk}
\end{figure}
\begin{figure}
\vspace*{4.5cm}
\special{hscale=35 vscale=35 hsize=1500 vsize=600
         hoffset=-20 voffset=160 angle=-90 psfile="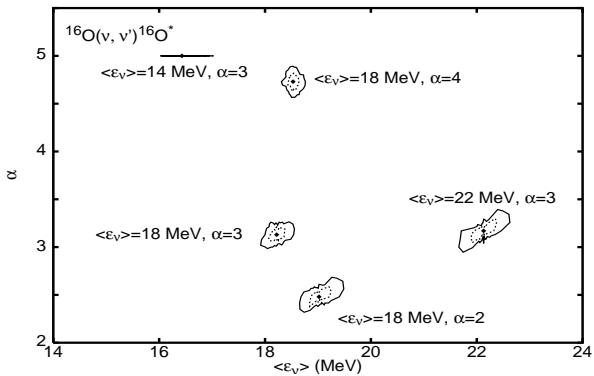"}
\caption{Reconstruction of the incident SN$\nu$ spectrum from cross sections folded with a synthetic spectrum (dots). The dotted and full lines represent the 90\% confidence levels for constructed spectra with 5 and 10\% uncertainties on the parameters $a^{\gamma_{i=1,5}}$.  
}
\label{noisplot}
\end{figure} 
Fig.~\ref{pbdif} 
illustrates that using the proposed technique, 
the  agreement between original and reconstructed response is sufficient to discriminate between spectra with only small separations in average energy. 
The right panel indicates that the method is able to distinguish between cross-section differences induced by more subtle variations in the spectra, such as slightly different widths. 

We now turn to our technique for extracting information about the SN neutrinos by combining the response in the
SN detector and information from low-energy beta-beams.
That this is not a trivial task, is demonstrated in Fig.~\ref{cumgelijk}, showing that different SN$\nu$ spectra can be extremely similar, e.g. when having  higher average energies but a reduced width. Obviously an efficient  tool 
 to distinguish between these spectra is needed.

Given the SN$\nu$ signal and a set of beta-beam response data, fitting the SN$\nu$ response in the detector to a linear combinations of beta-beam responses by minimizing $|\sigma^{fold}_{fit} - \sigma^{fold}_{signal}|$, will yield  the $N$ expansion parameters $a^{\gamma_i}$ and $\gamma$-values $\gamma^{i}$ that offer best agreement with the supernova data.  
Eq.~\ref{const} then immediately reveals the original SN$\nu$ energy-distribution.
The  
agreement in the predicted responses obtained for remarkably small values of $N$ guarantees the  efficiency of this formalism.
The accuracy of the technique is restricted mainly by detector limitations and is basically free of theoretical ambiguities. 

As a full Monte-Carlo analysis of supernova and beta-beam responses is as yet untimely, we illustrate that the proposed procedure can  successfully be inverted and that it is stable against experimental uncertainties, using a schematic model.  We create synthetic data represented by a cross section, folded with a synthetic neutrino energy distribution.
The original SN$\nu$ spectrum $n_{SN}$ that offers the best fit to these synthetic data is then determined by selecting the spectrum that minimizes the difference between a member of the $\sigma^{fold}_{SN}$ set and the synthetic response. 
The results of this process  can be seen 
in Fig.~\ref{noisplot}. In view of  the problems suggested by Fig.~\ref{cumgelijk}, the agreement is very good, especially at higher energies.
Experimental uncertainties in the beta-beam measurement and detector signal will give rise to noise on the expansion parameters that are determined 
from fitting the SN$\nu$ signal to a combination of beta-beam data.
To estimate the effect of this on the reconstruction of the SN$\nu$ spectrum, we repeated the   procedure for values of the coefficients $a_i$ which vary up to 10\%.  The mild scatter this produces, shows that the method is stable against uncertainties at this level.

In practice,  the signal in the detector will  be the superposition  of the $\nu_e$, $\overline{\nu}_e$ and the heavy-flavor spectra.  
Oscillations will  mix these original spectra.
 Though that leaves the NC signal  unchanged, the flavor shifts between heavy and electron (anti)neutrinos will most likely induce linear combinations of the undistorted  CC response. In terms of the proposed procedure  this means that one has to deal with an 'extra' linear combination superimposed on the one of Eq.~\ref{const}.  Comparing the actual combination observed  in NC and CC channels will provide information about the way the neutrinos were mixed by oscillations.

Summarizing, we studied the potential use of a low-energy beta-beam facility for the understanding of supernova physics.  We showed 
that using appropriate linear combinations of beta-beam spectra allows for a very accurate reproduction of  supernova-neutrino responses. 
The proposed procedure allows one to
make the most
 efficient use of the neutrino signal to filter information about the supernova
from the response in a terrestrial detector, thus showing that beta-beams are able to provide information about the spectra supernova-neutrinos were released with at the explosion site, in a direct way. 

\acknowledgments
The authors would like to thank C.~Volpe, M.~Lindroos,  and K.~Heyde for interesting discussions. N.J.~thanks the Fund for Scientific Research Flanders for financial support. G.C.M.~acknowledges support from the Department of Energy, under contract DE-FG02-02ER41216.

\end{document}